\documentclass[useAMS,usenatbib]{mn2e}

\usepackage{rotating}
\usepackage{multirow}

\usepackage{epsfig}
\usepackage{subfigure}

\usepackage{graphicx}
\DeclareGraphicsExtensions{.ps}

\makeatletter
\let\jnl@style=\rm
\def\ref@jnl#1{{\jnl@style#1}}

\def\aj{\ref@jnl{AJ}}                   
\def\araa{\ref@jnl{ARA\&A}}             
\def\apj{\ref@jnl{ApJ}}                 
\def\apjl{\ref@jnl{ApJ}}                
\def\apjs{\ref@jnl{ApJS}}               
\def\ao{\ref@jnl{Appl.~Opt.}}           
\def\apss{\ref@jnl{Ap\&SS}}             
\def\aap{\ref@jnl{A\&A}}                
\def\aapr{\ref@jnl{A\&A~Rev.}}          
\def\aaps{\ref@jnl{A\&AS}}              
\def\azh{\ref@jnl{AZh}}                 
\def\baas{\ref@jnl{BAAS}}               
\def\cjaa{\ref@jnl{ChJAA}}		
\def\jrasc{\ref@jnl{JRASC}}             
\def\memras{\ref@jnl{MmRAS}}            
\def\mnras{\ref@jnl{MNRAS}}             
\def\nar{\ref@jnl{NewAR}}               
\def\na{\ref@jnl{NewA}}                 
\def\pra{\ref@jnl{Phys.~Rev.~A}}        
\def\prb{\ref@jnl{Phys.~Rev.~B}}        
\def\prc{\ref@jnl{Phys.~Rev.~C}}        
\def\prd{\ref@jnl{Phys.~Rev.~D}}        
\def\pre{\ref@jnl{Phys.~Rev.~E}}        
\def\prl{\ref@jnl{Phys.~Rev.~Lett.}}    
\def\pasp{\ref@jnl{PASP}}               
\def\pasj{\ref@jnl{PASJ}}               
\def\qjras{\ref@jnl{QJRAS}}             
\def\skytel{\ref@jnl{S\&T}}             
\def\solphys{\ref@jnl{Sol.~Phys.}}      
\def\sovast{\ref@jnl{Soviet~Ast.}}      
\def\ssr{\ref@jnl{Space~Sci.~Rev.}}     
\def\zap{\ref@jnl{ZAp}}                 
\def\nat{\ref@jnl{Nature}}              
\def\iaucirc{\ref@jnl{IAU~Circ.}}       
\def\aplett{\ref@jnl{Astrophys.~Lett.}} 
\def\apspr{\ref@jnl{Astrophys.~Space~Phys.~Res.}}
\def\bain{\ref@jnl{Bull.~Astron.~Inst.~Netherlands}}
\def\fcp{\ref@jnl{Fund.~Cosmic~Phys.}}  
\def\gca{\ref@jnl{Geochim.~Cosmochim.~Acta}}   
\def\grl{\ref@jnl{Geophys.~Res.~Lett.}} 
\def\jcp{\ref@jnl{J.~Chem.~Phys.}}      
\def\jgr{\ref@jnl{J.~Geophys.~Res.}}    
\def\jqsrt{\ref@jnl{J.~Quant.~Spec.~Radiat.~Transf.}}
\def\memsai{\ref@jnl{Mem.~Soc.~Astron.~Italiana}}
\def\nphysa{\ref@jnl{Nucl.~Phys.~A}}   
\def\physrep{\ref@jnl{Phys.~Rep.}}   
\def\physscr{\ref@jnl{Phys.~Scr}}   
\def\planss{\ref@jnl{Planet.~Space~Sci.}}   
\def\procspie{\ref@jnl{Proc.~SPIE}}   

\begin{document}


\title{The 1.4 GHz radio properties of hard X-ray selected AGN}
 


\author[Francesca Panessa]{F. Panessa$^1$\thanks{E-mail: francesca.panessa@iaps.inaf.it},
A. Tarchi$^2$, P. Castangia$^2$, E. Maiorano$^3$, L. Bassani$^3$, G. Bicknell$^4$,
\newauthor A. Bazzano$^1$, A.J. Bird$^5$, A. Malizia$^3$, P. Ubertini$^1$ \\
$^1$ INAF - Istituto di Astrofisica e Planetologia Spaziali di Roma (IAPS), Via del Fosso del Cavaliere 100, 00133 Roma, Italy\\
$^2$ Osservatorio Astronomico di Cagliari (OAC-INAF),Via della Scienza 5,  09047 Selargius (CA), Italy \\ 
$^3$ Istituto di Astrofisica Spaziale e Fisica Cosmica (IASF-INAF), Via P. Gobetti 101, 40129 Bologna, Italy \\
$^4$ Research School of Astronomy \& Astrophysics, Mt Stromlo Observatory, Cotter Rd., Weston, ACT 2611, Australia\\
$^5$ Physics \& Astronomy, University of Southampton, Highfield, Southampton SO17 1BJ, U.K.\\
}
\date{}

\maketitle

\begin{abstract}

We have analyzed the NVSS and SUMSS data at 1.4 GHz and 843 MHz for a well defined complete sample of 
hard X-ray AGN observed by INTEGRAL. A large number (70/79) of sources are detected in the radio band,
showing a wide range of  radio morphologies, from unresolved or slightly resolved cores to extended emission over several hundreds
of kpc scales. The radio fluxes have been correlated with the 2-10 keV and 20-100 keV emission, revealing significant correlations
with slopes consistent with those expected for radiatively efficient accreting systems.
The high energy emission coming from the inner accretion regions correlates with the radio emission averaged over hundreds of kpc scales (i.e.,  thousands of years).
\end{abstract}


\section{Introduction}

How Active Galactic Nuclei (AGN) are involved in the process of galaxy formation
is one of the main issues in modern astrophysics. AGN and galaxy evolution are
connected through feedback processes regulating both the accretion history and the star-formation
(e.g., Cattaneo et al. 2009). The mechanical (kinetic) energy released by accelerated particles in jets
is one of the possible sources of AGN feedback into the interstellar matter of the host galaxy (Croton et al. 2006,
Cattaneo et al. 2006, Marulli et al. 2008). The knowledge of
the connection between the radio emission (linked to the jet) and the X-ray emission (from accretion) is one 
of the key ingredients to build up more robust AGN-galaxy evolutionary models (La Franca, Melini, Fiore et al. 2010,
Bonchi et al. 2013). 
The X-ray versus radio connection is also fundamental within the disk-jet coupling paradigm in accreting systems (Merloni, et al. 2003, Falcke et al. 2004).
However, the details of the link between the accreting disk, the X-ray corona and the jet
are still under investigation.

The relation between the radio and the X-ray emission has been explored for different classes of AGN, with
correlation slopes typically distributed around unity (Canosa et al. 1999, Brinkmann et al. 2000, Panessa et al. 2007).
The discovery of a ``fundamental plane" relation between the X-ray and radio luminosities, 
and the black hole mass (Merloni et al. 2003; Falcke et al. 2004), has been interpreted in the framework of a radiatively inefficient accretion flow accompanied by 
a relativistic jet at work both in X-ray Binaries (XRBs) and low luminosity AGN (LLAGN) (e.g., Narayan \& Yi 1994). 
Recently, Coriat et al. (2011) have shown that some XRBs follow a steeper radio-X-ray correlation (correlation slope of 1.4)
compared to the standard 0.5-0.7 (Gallo et al. 2003, Merloni et al. 2003), suggesting that these sources are accreting at high Eddington ratios
and are radiatively efficient. Dong et al. (2014) have confirmed this result for a sample of bright radiatively efficient AGN, with a positive
correlation slope of $\sim$ 1.6, justified within the disk-corona model for efficient accreting sources (e.g., Cao et al. 2014). 
On the other hand, Burlon et al. (2013) have explored the association between hard X-ray AGN observed by Swift/BAT
and AT20G radio sources in the southern sky, finding a possible correlation between the 15-55 keV and 20 GHz luminosities (slope of $\sim$ 0.7),
however likely driven by distances effects and therefore not intrinsically significant.

Here we aim at testing the validity of such correlations 
in a well defined complete sample of relatively high luminosity AGN, selected at hard X-ray as those observed by the INTEGRAL satellite (e.g., Bird et al. 2010).
INTEGRAL/IBIS is surveying the sky above 20 keV with a
sensitivity better than a few mCrab~\footnote{The conversion factor in the 20-40 keV range is: 10 mCrab $=$ 7.57 $\times$ 10$^{-11}$ erg cm$^{-2}$ s$^{-1}$, see Bird et al. (2010).} and a point source location accuracy of 1-3 arcmin depending on the source significance and the observation exposure
 (Bird et al. 2007). Hard X-ray selected samples are almost unbiased with respect
to absorption and are mostly located in the nearby Universe, reducing as much as possible the selection effects (Malizia et al. 2009).
The sample used here comprises bright AGN (41.5 $<$ L$_{2-10 keV}$ $<$ 44.5 in erg/sec), with Eddington ratios L$_{bol}$/L$_{Edd}$ $\geq$ 10$^{-3}$, 
good candidates for radiating at efficient accretion regimes. 

In section 2 we present the INTEGRAL sample and the treatment of the X-ray data; in section 3 and 4 we describe the radio data analysis and the radio properties of the sample;
in section 5 the radio versus X-ray correlation statistical results are presented; in section 6 and 7 we report our discussion and conclusions.
Throughout this paper we assume a flat $\Lambda$CDM cosmology with ($\Omega_{\rm M}$,  
$\Omega_{\rm\Lambda}) = (0.3$,0.7) and a Hubble constant of 70 km s$^{-1}$ Mpc$^{-1}$ (Jarosik et al. 2011).

\section{The sample and the X-ray data}

\begin{figure*}
\begin{center}
\includegraphics[width=0.4\textwidth,height=0.3\textheight,angle=0]{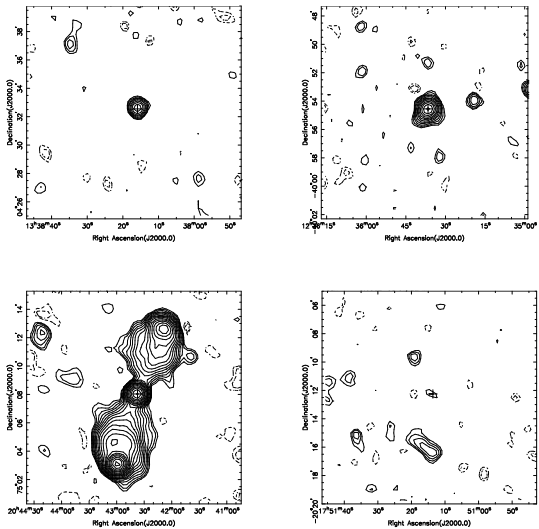}
\caption{Prototype of the different morphological classes proposed, i.e., top left: NGC~5252 (U), top right: NGC~4507 (S), bottom left: 4C~74.26 (R) 
and bottom right: IGR~J17513-2011 (A). The NED optical position is marked with a cross.}
\label{morf}
\end{center}
\end{figure*}

The sample is extracted from the third INTEGRAL/IBIS survey which lists 
around 140 (identified and candidates) AGN (Bird et al. 2007). To this large sample, 
the V/V$_{max}$ test~\footnote{Introduced by Schmidt (1968),
is a test of the uniformity of the space distribution in a flux-limited sample. The test compares the volume (V) enclosed within the distance of a source with the maximum volume (V$_{max}$) 
that would be enclosed at the maximum redshift at which the object would be detectable.} 
has been applied to obtain a complete sample of 79 AGN selected in
the hard (20-40 keV) X-ray band above $\sim$ 5 sigma confidence level
(see Malizia et al. 2009 for the detailed definition of the method and sample); the sample includes 46 Seyfert 1 (including 6 Narrow Line objects) and 33 Seyfert 2. 
The original sample in Malizia et al.  (2009) contains also 9 blazars, which we have excluded in the present study 
because Doppler boosting effects prevent from a correct determination of the intrinsic radio luminosity.  

The X-ray properties of the sample are described in Malizia et al. (2009) where the 20-100 keV fluxes were taken from the INTEGRAL 
survey (Bird et al. 2007), while the 2-10 keV observed fluxes and column densities 
were collected from the literature. In order to obtain the 2-10 keV unabsorbed fluxes from the observed ones, 
we have here applied a correction factor proportional to the measured X-ray column density N$_{H}$ 
as in column 8 of table 1 in Malizia et al. (2009), assuming simple photoelectric absorption in XSPEC.
The sample includes five Compton thick AGN (MKN~3, NGC~3281, NGC~4945, Circinus galaxy, IGR~J16351-5806 and NGC~1068), for which both 2-10 keV and 20-100 keV fluxes require a correction
(the hard X-ray fluxes are very marginally affected by absorption for N$_{H}$ $<$ 10$^{24}$ cm$^{-2}$).
In Malizia et al. (2009) the correction in the 20-40 keV flux as a function of N$_H$ was calculated using a simple absorbed power-law model in XSPEC, 
obtaining an average flux reduction negligible below log N$_H$=24 and progressively more important thereafter 
(8\%, 25\% and 64\% reduction in the ranges 24-24.5, 24.5-25, and 25-25.5 respectively). 
In Burlon et al. (2011) the observed/intrinsic flux ratio is plotted against the N$_H$ value, a few Compton thick sources
with available observed and intrinsic luminosities are overplotted to the predicted trend for local AGN.
The flux reduction is 40\%, 85\% and 94\% in the ranges 24-24.5, 24.5-25, and 25-25.5 respectively. 
In this work we have applied a tentative flux correction as in Burlon et al. (2011).
We should also consider that for the heavily absorbed source NGC~1068 (N$_{H}$ $>$ 10$^{25}$ cm$^{-2}$), 
an estimate of the intrinsic hard luminosity is difficult to obtain, given its spectral complexity and the unknown precise value of N$_{H}$
(Iwasawa et al. 1997, Matt et al. 2004).

\section{The radio data}

The NRAO VLA Sky Survey survey (NVSS; Condon et al. 1998) is a 1.4 GHz continuum survey covering the entire sky north of --40 deg declination. The restoring beam of the NVSS maps is 
45$\arcsec$ full width at half-maximum (FWHM) and the rms brightness fluctuations in the images are about 0.45 mJy/beam.  
The Sydney University Molonglo Sky Survey (SUMSS, Bock et al. 1999, Mauch et al. 2003) survey at 843 MHz covers the whole sky 
south of declination --30 deg, and produces images with a resolution of 45$\arcsec$ $\times$ 45$\arcsec$ cosec \textbar dec\textbar  \
and an rms noise level of $\sim$1 mJy/beam.
Thus, SUMSS and NVSS have similar sensitivity and resolution, and together cover the whole sky. 

We extraxted maps of about 30$\arcmin$$\times$30$\arcmin$ in size from the Postage Stamp Services available for both surveys 
centering the requested map at the coordinates of the optical 
nucleus of the galaxy associated by Malizia et al. (2009) to the 
INTEGRAL source.
FITS images were imported into NRAO Astronomical Image Processing System (AIPS) and analyzed using standard AIPS tasks. 
When the radio sources are either unresolved or slightly resolved, 
all parameters and associated errors were determined through a 
Gaussian fit of each component using the AIPS task IMFIT. 
For extended sources, coordinates and flux density of the radio 
emission peaks were obtained using the AIPS task TVSTAT. The integrated flux density 
 were also obtained using TVSTAT over a polygon encompassing the radio emission visible down to 
 the 3$\mathrm{\sigma}$ level of the map to avoid confusion with the extended 
background emission.

The measured quantities are collected in table 1.

For unresolved sources, i.e. those for which IMFIT computes FWHM values smaller
than one-half the restoring beam\footnote{The restoring beam size of
the NVSS and SUMSS images is  45$\arcsec$ $\times$ 45$\arcsec$ and  43$\arcsec$ $\times$ 43$\arcsec$, respectively.} size in both dimensions, the reported sizes are set to one-half the restoring beam size at FWHM. 
These dimensions are obviously upper limits because the true 
dimensions are not known. 
We define slightly resolved sources as those for which IMFIT computes values for the size larger than one-half the restoring beam size at FWHM in at least one of the 
two dimensions. In this case the reported angular size is the deconvolved size provided by IMFIT for the slightly-resolved dimension(s) and one-half the restoring beam size at FWHM, otherwise. For those sources marked as resolved, viz. those for which fitting the emission with a single Gaussian 
was clearly not suitable, the extension of the radio source has been 
computed directly from the map by measuring the extension of the radio emission visible above a 3$\mathrm{\sigma}$ level in two orthogonal directions.  
Note that the different frequency between of the two surveys does not allow a direct comparison of the linear dimensions, different radio structures are indeed detected
depending on the observational frequency. 
The errors on the estimated quantities were directly provided by the 
task IMFIT (when used). For the extended sources we decided 
to use the noise of the map (determined from a source-free rectangular region) as the uncertainty on the peak flux density, while the uncertainty on the integrated flux density was computed using 
the formula:
$$
\Delta S_i= \sigma \times \sqrt[2]{\frac{pix_{area}}{pix_{beam}}}
 $$
where $\sigma$ is the r.m.s noise,  pix$_{area}$ and pix$_{beam}$ are the number of pixel within the area over which we integrate and the pixel in the beam, respectively.

For the SUMSS sources, the flux density at 843~MHz has been converted into
1.4~GHz flux density assuming S$_{\nu}$ $\propto$ $\nu^{- \alpha}$ with ${\alpha}$ = 0.7.
This is an average slope value expected for radio optically thin synchrotron emission,
however slopes can vary from steep to flat or inverted depending on the source, usually Seyfert radio cores show
flat/inverted slopes on arcsecond scales (Ho \& Ulvestad 2001),
while steep optically thin synchrotron slopes are associated to extended structures such as lobes.
The flux measured can vary by $\sim$ 30\% from steep to flat/inverted slopes during the conversion from 843 MHz to 1.4 GHz,
however this affects only marginally the results on our correlations (see next Section) as 
the SUMSS sources will probably have spectral slope variously distributed, and these differences
will be averaged in the statistical treatment of the fluxes.

The absolute calibration of both the NVSS and SUMSS flux density scale is 3\% (Kaplan et al. 1998, 
uncertainties relative to the Baars et al. 1977 scale).

Following the classification outlined before and adopting definitions consistent with those of Ulvestad  \& Wilson (1984) and Ho \& Ulvestad (2000) for the radio morphology we distinguish between four classes in our sample: ``U'' (single, unresolved), ``S'' (single, slightly resolved), ``R'' (diffuse or linear, clearly resolved), and ``A'' (ambiguous, marginally detected).
By way of example, in figure~\ref{morf} we show radio images of four galaxies selected as prototype of the different morphological classes proposed, i.e., NGC~5252, NGC~4507, 4C~74.26 and IGR~J17513-2011, as U, S, R and A, respectively.

Of the two sources labeled as ``ambiguous'', one is only marginally detected (IGR~J17513-2011) and one is possibly contaminated by a background radio source (IGR~J16185-5928).
Note that also in the NVSS image of IGR~J16385-2057, there is a background radio galaxy which might contaminate the radio emission measured, however this source is still under study and we report it as S (see Tarchi et al. in preparation).

\begin{table*}
\begin{center}
\centerline{Table 1: The NVSS data for the INTEGRAL/IBIS complete sample of AGN}
 \begin{tabular}{l r r r c r r r r r r l l l}
  \hline\hline
Name              &      Class   &       $z$   & L$_{X}$ & L$_{HX}$ & S$_{p}$  & S$_{i}$ & L$_{p}$ & L$_{i}$ & D$_{min}$ & 
D$_{maj}$ & Mor.  & M & Ref.\\
(1)              &      (2)     &  (3) & (4) & (5) & (6) & (7) & (8) & (9) & (10) & (11) & (12) & (13) & (14) \\
\hline

IGR J00333+6122    &   Sy 1.5  &   0.1050    & 44.19	   & 	44.44	& 	  9.0	&	 9.5   &     39.46   &     39.48   &	$<$ 49.09  &	$<$ 49.09 &   U 	 &   8.5 &  1	    \\ 
NGC 788            &   Sy 2    &   0.0136    & 42.86	   & 	43.29	& 	  6.0	&	19.5   &     37.61   &     38.17   &	    27.13  &	     35.6 &   R 	 &   7.5 &  2	    \\ 
NGC 1068           &   Sy 2    &   0.0038    & 42.95	   & 	43.35	&      4.5e3	&     5.0e3    &     39.27   &     39.32   &	    17.05  &	    18.95 &   R 	 &   7.2 &  2	    \\ 
QSO B0241+62       &   Sy 1    &   0.0440    & 43.87	   & 	44.41	& 	353.8	&      374.9   &     40.30   &     40.32   &	$<$ 20.57  &	$<$ 20.57 &   U 	 &   8.5 &  3	   \\ 
NGC 1142           &   Sy 2    &   0.0288    & 43.82	   & 	43.96	& 	139.5	&      158.4   &     39.53   &     39.58   &	$<$ 13.46  &	$<$ 13.46 &   U 	 &   9.4 &  2	    \\ 
B3 0309+411B       &   Sy 1    &   0.1360    & 45.04	   & 	44.99	& 	344.8	&      389.5   &     41.27   &     41.44   &	    423.9  &	     1373 &   R 	 &   $-$ &  $-$    \\ 
NGC 1275           &   Sy 2    &   0.0175    & 42.89	   & 	43.38	&     2.3e4	&    2.3e4     &     41.30   &     41.33   &	    93.81  &	    109.1 &   R 	 &   8.5 &  2	    \\ 
3C 111             &   Sy 1    &   0.0485    & 44.52	   & 	44.66	&      3.5	&   15.1	&    41.38   &     42.01    &	   320.5   &	     393  &   R	   	 &   9.6 &  2	  \\ 
LEDA 168563        &   Sy 1    &   0.0290    & 43.90	   & 	44.04	& 	 12.4	&	15.0   &     38.48   &     38.56   &	$<$ 13.56  &        15.68 &   S 	 &   8.0 &  4	  \\ 
4U 0517+17         &   Sy 1.5  &   0.0179    & 43.23	   & 	43.62	& 	  6.0	&	 6.1   &     37.75   &     37.76   &	$<$ 8.368  &        8.484 &   S 	 &   7.0 &  5	   \\ 
MCG+08-11-11       &   Sy 1.5  &   0.0205    & 43.69	   & 	43.74	& 	228.5	&      249.3   &     39.45   &     39.50   &	    38.33  &	    84.34 &   R 	 &   8.1 &  2	    \\ 
Mkn 3              &   Sy 2    &   0.0135    & 44.17	   & 	43.79	&      1.1e3	&     1.1e3    &     39.76   &     39.77   &	$<$ 6.311  &	$<$ 6.311 &   U 	 &   8.7 &  2	    \\ 
Mkn 6	           &   Sy 1.5  &   0.0188    & 43.27	   & 	43.57	& 	259.6	&      281.2   &     39.43   &     39.46   &	$<$ 8.789  &	$<$ 8.789 &   U 	 &   8.2 &  2	    \\ 
IGR J07565-4139    &   Sy 2    &   0.0210    & 42.14	   & 	43.21	& 	  5.4	&	 3.9   &     37.84   &     37.70   &	$<$ 9.381  &	    14.05 &   S$^{*}$ 	 &   $-$ &  $-$   \\ 
IGR J07597-3842    &   Sy 1.2  &   0.0400    & 43.90	   & 	44.11	& 	  3.8	&	 3.6   &     38.24   &     38.22   &	$<$  18.7  &	$<$  18.7 &   U 	 &   8.3 &  1	   \\ 
ESO 209-12         &   Sy 1    &   0.0396    & 43.43	   & 	43.87	& 	 37.6	&	45.6   &     39.23   &     39.32   &	$<$ 17.69  &	     36.7 &   S$^{*}$ 	 &   $-$ &  $-$   \\ 
FRL 1146           &   Sy 1.5  &   0.0316    & 43.39	   & 	43.56	& 	 13.1	&	21.8   &     38.61   &     38.82   &	    63.03  &	    94.55 &   R 	 &   $-$ &  $-$    \\ 
SWIFT J0917.2-6221 &   Sy 1    &   0.0573    & 44.17	   & 	44.17	& 	  41.4   &	 46.1	&    39.60   &     39.64    &	$<$  25.6   &	   35.35  &   S$^{*}$	 &   9.9 &  6	    \\ 
MCG-05-23-16       &   Sy 2    &   0.0085    & 43.09	   & 	43.34	& 	 14.1	&	14.8   &     37.47   &     37.49   &	$<$ 3.974  &	$<$ 3.974 &   U 	 &   6.3 &  2	   \\ 
IGR J09523-6231    &   Sy 1.9  &   0.2520    & 45.28	   & 	45.22	& 	  5.6	&	 5.0   &     40.02   &     39.97   &	$<$ 112.6  &	$<$ 112.6 &   U$^{*}$ 	 &   7.6 &  2	   \\ 
SWIFT J1009.3-4250 &   Sy 2	&  0.0330    & 42.96	   & 	43.81	& 	  17.2   &	 16.8	&    38.74   &     38.73    &	$<$ 14.74   &	   28.52  &   S$^{*}$	 &   $-$ &  $-$    \\ 
NGC 3281           &   Sy 2    &   0.0115    & 43.68	   & 	43.35	& 	 73.1	&	81.6   &     38.45   &     38.50   &	$<$ 5.376  &	$<$ 5.376 &   U 	 &   8.0 &  2	    \\ 
SWIFT J1038.8-4942 &   Sy 1.5	&  0.0600    & 44.04	   & 	44.04	& 	   3.1   &	  2.9	&    38.51   &     38.48    &	$<$  26.8   &	   30.67  &   S$^{*}$	 &   $-$ &  $-$    \\ 
IGR J10404-4625    &   Sy 2    &   0.2392    & 44.93	   & 	45.58	& 	 42.0	&	43.3   &     40.84   &     40.86   &	$<$ 106.9  &	      204 &   S$^{*}$ 	 &   $-$ &  $-$   \\ 
NGC 3783           &   Sy 1    &   0.0097    & 43.07	   & 	43.42	& 	 35.5	&	45.5   &     38.00   &     38.15   &	    16.93  &	    55.63 &   R 	 &   7.5 &  2	   \\ 
IGR J12026-5349    &   Sy 2    &   0.0280    & 43.11	   & 	43.81	& 	 49.9	&	57.5   &     39.06   &     39.34   &	    87.27  &	    143.1 &   R 	 &   $-$ &  $-$    \\ 
NGC 4151           &   Sy 1.5	&  0.0033    & 43.05	   & 	43.16	& 	348.9	&      370.7   &     38.04   &     38.07    &	$<$ 1.543  &	$<$ 1.543 &   U	   	 &   7.5 &  2	  \\ 
Mkn 50             &   Sy 1    &   0.0234    & 43.05	   & 	43.20	&    $<$  1.5	&   $-$        &  $<$ 37.37   &     $-$    &	    $-$    &          $-$ &  $-$	 &   7.5 &  7	  \\ 
NGC 4388           &   Sy 2    &   0.0084    & 43.00	   & 	43.54	& 	 93.5	&      125.5   &     38.29   &     38.48   &	    28.27  &	    49.22 &   R 	 &   7.2 &  2	    \\ 
NGC 4507           &   Sy 2    &   0.0118    & 43.08	   & 	43.70	& 	 47.6	&	67.1   &     38.28   &     38.43   &	$<$ 5.999  &	    8.086 &   R 	 &   7.6 &  2	    \\ 
LEDA 170194        &   Sy 2    &   0.0360    & 43.48	   & 	44.22	& 	 35.4	&	40.6   &     39.13   &     39.23   &	    76.3   &	    130.2 &   R 	 &   8.9 &  2	    \\ 
NGC 4593           &   Sy 1    &   0.0090    & 42.80	   & 	43.09	& 	  4.0	&	 8.3   &     37.18   &     37.81   &	    15.71  &	     35.9 &   R 	 &   7.0 &  2	   \\ 
IGR J12415-5750    &   Sy 1    &   0.0230    & 42.93	   & 	43.40	& 	 15.1	&	16.9   &     38.37   &     38.42   &	$<$ 10.27  &	    15.83 &   S$^{*}$ 	 &   8.0 &  1	  \\ 
NGC 4945           &   Sy 2    &   0.0019    & 42.20	   & 	42.55	&      3.9e3	&     4.5e3    &     38.62   &     38.89   &	    10.19  &	       27 &   R 	 &   6.2 &  2	    \\ 
IGR J13091+1137    &   Sy 2    &   0.0251    & 43.44	   & 	43.77	&    $<$  1.4	&   $-$        &  $<$ 37.42   &     $-$     &	    $-$    &          $-$ &  $-$	 &   8.6 &  2		   \\ 
IGR J13109-5552    &   Sy 1    &   0.0850    & 43.88	   & 	44.56	& 	408.9	&      530.5   &     40.94   &     41.21   &	    201.3  &	    667.6 &   R$^{*}$ 	 &   $-$ &  $-$    \\ 
Cen A              &   Sy 2    &   0.0018    & 42.36	   & 	42.70	&      6.5e3	&    2.1e4     &     39.35   &     40.53   &	    4.264  &	    33.66 &   R$^{*}$ 	 &   8.0 &  2	    \\ 
MCG-06-30-15       &   Sy 1.2  &   0.0077    & 42.63	   & 	42.76	&    $<$  1.5	&   $-$        &  $<$ 36.40   &     $-$     &	    $-$    &          $-$ &  $-$	 &   6.7 &  2		   \\ 
NGC 5252           &   Sy 2    &   0.0230    & 43.72	   & 	43.71	& 	 15.9	&	16.9   &     38.39   &     38.41   &	$<$ 10.75  &	$<$ 10.75 &   U 	 &   9.0 &  2	   \\ 
4U 1344-60         &   Sy 1.5  &   0.0130    & 43.70	   & 	43.41	& 	  $<$ 9.2  &	$-$   &      37.50   &     41.54   &	    $-$    &          $-$ &  $-$$^{*}$	 &   $-$ &  $-$ 	  \\ 
IC 4329A           &   Sy 1.2  &   0.0160    & 43.74	   & 	44.06	& 	 63.6	&	69.0   &     38.67   &     38.76   &	     37.9  &	    113.7 &   R 	 &   7.0 &  2	    \\ 
Circinus Galaxy    &   Sy 2    &   0.0014    & 42.39	   & 	42.15	& 	715.3	&     1256.9   &     37.64   &     38.04   &	     10.3  &	    16.41 &   R$^{*}$ 	 &   6.0 &  2	   \\ 
NGC 5506           &   Sy 1.9  &   0.0062    & 42.83	   & 	42.85	& 	332.3	&      345.7   &     38.57   &     38.59   &	$<$ 2.898  &	$<$ 2.898 &   U 	 &   6.7 &  2	    \\ 
ESO 511-G030       &   Sy 1    &   0.2239    & 45.13	   & 	45.56	& 	  8.3	&	14.0   &     40.08   &     40.31   &	    118.6  &	    224.8 &   R 	 &   8.7 &  8	   \\ 
IGR J14515-5542    &   Sy 2    &   0.0180    & 42.68	   & 	43.04	& 	 16.5	&	16.5   &     38.19   &     38.19   &	$<$ 8.041  &	    11.79 &   S$^{*}$ 	 &   $-$ &  $-$   \\ 
IC 4518A           &   Sy 2    &   0.0163    & 42.83	   & 	43.09	& 	144.3	&      184.2   &     39.05   &     39.33   &	    48.77  &	    71.12 &   R$^{*}$ 	 &   7.5 &  9     \\ 
IGR J16024-6107    &   Sy 2    &   0.0110    & 41.66	   & 	42.46	& 	  8.8	&	 6.1   &     37.49   &     37.33   &	$<$ 4.914  &	$<$ 4.914 &   U$^{*}$ 	 &   $-$ &  $-$    \\ 
IGR J16119-6036    &   Sy 1    &   0.0160    & 42.24	   & 	43.13	&  $<$    4.9	&	 $-$   &  $<$ 37.41   &     $-$     &       $-$    &          $-$ &  $-$$^{*}$	 &   $-$ &  $-$ 	  \\ 
IGR J16185-5928    &   NLS1    &   0.0350    & 43.10	   & 	43.67	& 	  4.7	&	 5.3   &     38.23   &     38.27   &	$<$ 15.64  &	    38.04 &   A$^{*}$	 &   7.4 &  1	    \\ 
IGR J16351-5806    &   Sy 2    &   0.0091    & 42.51	   & 	42.74	& 	 69.8	&	84.2   &     38.22   &     38.41   &	    27.23  &	    30.63 &   R$^{*}$ 	 &   $-$ &  $-$    \\ 
IGR J16385-2057    &   NLS1    &   0.0269    & 43.02	   & 	43.39	& 	  6.5	&	 7.2   &     38.13   &     38.24   &	$<$ 12.58  &	    15.34 &   S 	 &   6.8 &  1	   \\ 
IGR J16426+6536    &   NLS1    &   0.3230    & $-$	   & 	45.97	&     $<$ 1.2	&   $-$        &  $<$ 39.57   &     $-$     &	    $-$    &          $-$ &   $-$	 &   7   &  1		   \\ 
IGR J16482-3036    &   Sy 1    &   0.0310    & 43.60	   & 	43.79	& 	  2.4	&	 7.8   &     37.89   &     38.40   &	    57.97  &	    112.1 &   R 	 &   8.2 &  1	   \\ 
IGR J16558-5203    &   Sy 1.2  &   0.0540    & 44.33	   & 	44.31	& 	  5.3	&	 3.0   &     38.65   &     38.40   &	$<$ 24.12  &	$<$ 24.12 &   U$^{*}$ 	 &   7.9 &  1	   \\ 
NGC 6300           &   Sy 2    &   0.0037    & 41.87	   & 	42.28	& 	 18.8	&	64.3   &     36.92    &    37.54   &	    11.53  &	     20.3 &   R$^{*}$ 	 &   5.5 &  2	    \\ 
GRS 1734-292       &   Sy 1    &   0.0214    & 43.56	   & 	43.90	& 	 46.3	&	48.5   &     38.79    &    38.81   &	$<$    10  &	$<$    10 &   U 	 &   8.9 &  2	    \\ 
2E 1739.1-1210     &   Sy 1    &   0.0370    & 43.56	   & 	43.89	& 	  3.4	&	 3.1   &     38.13    &    38.09   &	$<$  17.3  &	$<$  17.3 &   U 	 &   8.2 &  1	  \\ 
\end{tabular}				        							      										   
\end{center}				        							      										  
\small					        							      										  
\end{table*}				        							      										  
					        							      										     				
\begin{table*}												      										  
\begin{center}												      										  
\centerline{Table 1: The NVSS data for the INTEGRAL/IBIS complete sample of AGN}			      										  
 \begin{tabular}{l r r r c r r  r r r r l l l}								      										  
  \hline\hline												      										  
Name             &      Class   &       $z$   & L$_{2-10}$ & L$_{20-100}$ &  S$_{p}$  & S$_{i}$ & L$_{p}$ & L$ _{i}$ & D$_{min}$
&  D$_{maj}$ & Mor. &  M  &Ref.\\			  
(1)              &      (2)     &  (3) & (4) & (5) & (6) & (7) & (8) & (9) & (10) & (11) & (12) & (13) & (14)\\	      									  
\hline													      										  
IGR J17488-3253     &   Sy 1  &   0.0200    & 43.07  &        43.52   &   $<$  1.2   &   $-$	    &	 $<$ 37.14   &    $-$  &         $-$    &        $-$ &  $-$	 &   $-$ & $-$     \\  
IGR J17513-2011     &   Sy 1.9&   0.0470    & 43.40  &        44.10   &        1.5   &        1.8   &	 37.99   &    38.06  &   $<$  21.97   &        59.68 &   A     	 &   6.0 & 2	      \\ 
IGR J18027-1455     &   Sy 1  &   0.0350    & 43.23  &        44.04   &        5.7   &        8.4   &	 38.51   &    38.87  &        30.54   &	         144 &   R	 &   $-$ & $-$      \\ 
IGR J18249-3243     &   Sy 1  &   0.3550    & 45.13  &        45.47   &     2.7e3    &     4.1e3    &	 43.01   &    43.18  &         1593   &	        1770 &   R	 &   $-$ & $-$      \\ 
IGR J18259-0706     &   Sy 1  &   0.0370    & 43.19  &        43.66   &        2.7   &        9.6   &	 38.66   &    38.89  &        69.19   &	       133.8 &   R	 &   $-$ & $-$      \\ 
ESO 103-35          &   Sy 2  &   0.0133    & 43.26  &        43.49   &       24.5   &       26.0   &	 38.10   &    38.13  &   $<$  5.941   &   $<$  5.941 &   U$^{*}$ &   7.1 & 2		   \\ 
3C 390.3            &   Sy 1  &   0.0561    & 44.34  &        44.60   &     4.2e3    &     7.0e3    &	 41.61   &    42.02  &        230.8   &        419.6 &   R	 &   8.5 & 2	  \\ 
2E 1853.7+1534      &   Sy 1  &   0.0840    & 44.25  &        44.56   &        3.5   &        3.4   &	 38.86   &    38.84  &   $<$  39.27   &   $<$  39.27 &   U	 &   8.2 & 1	 \\ 
IGR J19378-0617     &   NLS1  &   0.0106    & 42.91  &        42.64   &       39.3   &       44.0   &	 38.11   &    38.16  &   $<$  4.955   &   $<$  4.955 &   U	 &   6.8 & 10	   \\ 
NGC 6814            &   Sy 1.5&   0.0052    & 40.98  &        42.52   &       13.7   &       52.9   &	 37.07   &    37.66  &         18.8   &	       27.88 &   R	 &   7.1 & 2	     \\ 
Cyg A               &   Sy 2  &   0.0561    & 44.68  &        44.73   &   4.6e5      &      1.7e6   &	 43.81   &    44.17  &        279.8   &	       468.6 &   R	 &   9.4 & 11	    \\ 
IGR J20186+4043     &   Sy 2  &   0.0144    & 42.57  &        42.98   &  $<$   4.1   &   $-$	    &$<$ 37.39   &    $-$    &         $-$    &          $-$ &  $-$	 &   $-$ & $-$  	 \\ 
4C 74.26            &   Sy 1  &   0.1040    & 44.75  &        45.04   &      190.5   &      208.8   &	 41.25   &    41.76  &        259.3   &         1724 &   R	 &   9.6 & 2	     \\ 
S52116+81           &   Sy 1  &   0.0840    & 44.25  &        44.77   &      243.8   &      311.6   &	 40.71   &    40.99  &        345.6   &        701.6 &   R	 &   8.8 & 2	     \\ 
IGR J21247+5058     &   Sy 1  &   0.0200    & 43.81  &        43.96   &      210.6   &      399.6   &	 39.51   &    40.47  &        82.28   &        279.3 &   R	 &   6.6 & 1	     \\ 
SWIFT J2127.4+5654  &   NLS1  &   0.0140    & 42.89  &        43.04   &        6.0   &        7.3   &	 37.53   &    37.62  &   $<$  6.545   &        7.918 &   S	 &   7.2 & 1	 \\ 
RX J2135.9+4728     &   Sy 1 &   0.0250    & 43.00  &        43.32   &        6.6   &        7.2   &	 38.08   &    38.12  &   $<$  11.69   &   $<$  11.69 &   U	 &   $-$ & $-$    \\ 
NGC 7172            &   Sy 2  &   0.0087    & 42.51  &        43.11   &       33.3   &       39.1   &	 37.87   &    38.01  &        17.35   &        36.88 &   R	 &   7.7 & 2	   \\ 
MR 2251-178         &   Sy 1  &   0.0640    & 44.23  &        44.79   &       15.2   &       17.4   &	 39.26   &    39.32  &   $<$  29.92   &   $<$  29.92 &   U	 &   6.9 & 2	  \\ 
MCG-02-58-22        &   Sy 1.5&   0.0469    & 44.16  &        44.25   &       30.8   &       34.8   &	 39.29   &    39.35  &   $<$  21.93   &   $<$  21.93 &   U	 &   7.1 & 2	   \\ 
IGR J23308+7120     &   Sy 2  &   0.037     & 42.80  &        43.55   &        2.1   &        4.0   &	 37.92   &    38.20  &    	 25   &        40.31 &   S	 &   $-$ & $-$   \\ 
IGR J23524+5842     &   Sy 2  &   0.1640    & 44.41  &        44.85   &   $<$  1.5   &   $-$	    &$<$ 39.06   &    $-$    & 	       $-$    &          $-$ &  $-$	 &  $-$  & $-$  	 \\ 
\hline																						  
\end{tabular}																					  
\end{center}																					  
\small																						  
Notes:  Col. (1) Galaxy name; col. (2-3) optical classification and redshift from Malizia et al. (2009);  col. (4-5); 2-10 keV and 20-100 keV logarithmic luminosities in erg/sec, 
corrected for absorption; (6-7) peak flux density (mJy/beam) and integrated flux density (mJy) at 1.4 GHz; for the SUMSS data 							  
the 843 MHz flux has been converted to 1.4 GHz assuming a spectral index of 0.7; col. (8-9) 1.4 GHz peak and integrated logarithmic luminosities in				  
erg/s; col. (10-11) Minor and major axis sizes in kpc measured from the deconvolved sizes at 1.4 GHz from the NVSS and at 843 MHz from the SUMSS; col. (12): 			  
radio morphology class, adopting the definitions by Ulvestad \& Wilson (1984): "U" (single, unresolved), "S" (single, slightly resolved), "R" (resolved) and "A" (ambiguous).
$^{*}$ Source from the SUMSS survey; col. (13-14): logarithm of black hole mass (in M$_{\odot}$) and relative reference: 1) Masetti web page: http://www.iasfbo.inaf.it/$\sim$masetti/IGR/main.html,
2) Wang et al. 2009, 3) Beckmann et al. 2009, 4) Raimundo et al. 2012, 5) Stalin et al. 2011, 6) Perez et al. 1989, 
7) Barth et al. 2011, 8) Winter et al. 2009, 9) Kim et al. 2010, 10) Alonso-Herrero et al. 2013,
11) McKernan et al. 2010.
 \end{table*}

\section{The radio properties of the INTEGRAL sample}																  
																						  
The complete hard X-ray INTEGRAL sample offers the opportunity to study the characteristics of											  
the radio emission at kpc scales for a sample of bright local AGN.														  
 																						  
A galaxy was considered detected in the NVSS/SUMSS maps when a source with a peak flux density greater or equal than three times the rms noise of the map was found coincident, within the uncertainties, with the optical position given in Malizia et al. (2009). Of the 79 sources in this sample, 58 sources belong to the NVSS survey, the remaining 21 have been observed with the SUMSS survey.
In our sample, only 9 sources have not been detected. We have checked for a possible detection in other observations available in literature and,				  
of the nine undetected sources, only MCG-6-30-15 has been detected with VLA at 8.4 GHz with 0.64 mJy (Mundell et al. 2009), while IGR~J17488-3253 has an upper limit of 2.1 mJy with GMRT at 0.61 GHz (Pandey et al. 2006), IGR~J20186+4043 has been observed with VLA-D and has been associated to an extended radio source, possibility the AGN counterpart, although the association is not secure (Bykov et al. 2006). The remaining sources have no other radio data.
																						  
Limiting this work to the NVSS and SUMSS observations, the sample detection rate is therefore 89\%. This fraction is similar to the 85\% found for an optically selected sample observed with VLA (1.4 and 5 GHz) at high resolution ($\sim$1$\arcsec$) by Ho \& Ulvestad (2001). The first and only (up to now) radio follow-up of hard X-ray selected sources has been reported in Burlon et al. (2013),
where only 20\% of the Swift/BAT AGN are associated with a 20 GHz AT20G detection in the southern sky. However, the low detection rate could be ascribed			  
to the low sensitivity of the AT20G survey ($\sim$40 mJy) and to the higher observing frequency with respect to the NVSS/SUMSS surveys, suggesting that only bright radio sources are revealed.
Indeed, the radio luminosities of Seyfert galaxies in Burlon et al. (2013) are almost uniformly distributed from 10$^{39}$ - 10$^{44}$ erg/sec, while INTEGRAL sources 
show 1.4 GHz peak luminosities that vary from $\sim$ 10$^{37}$ erg/sec up to $\sim$ 10$^{44}$ erg/sec, with a peak distribution around 
10$^{37}$ - 10$^{40}$ erg/sec.
However, INTEGRAL AGN are still radio bright when compared with
optically selected Seyferts, whose radio luminosities are typically found in the range 10$^{34-42}$ erg/sec (Panessa et al. 2007, Nagar et al. 2002). 

\begin{figure*}
\begin{center}
\parbox{16cm}{
\includegraphics[width=0.4\textwidth,height=0.3\textheight,angle=0]{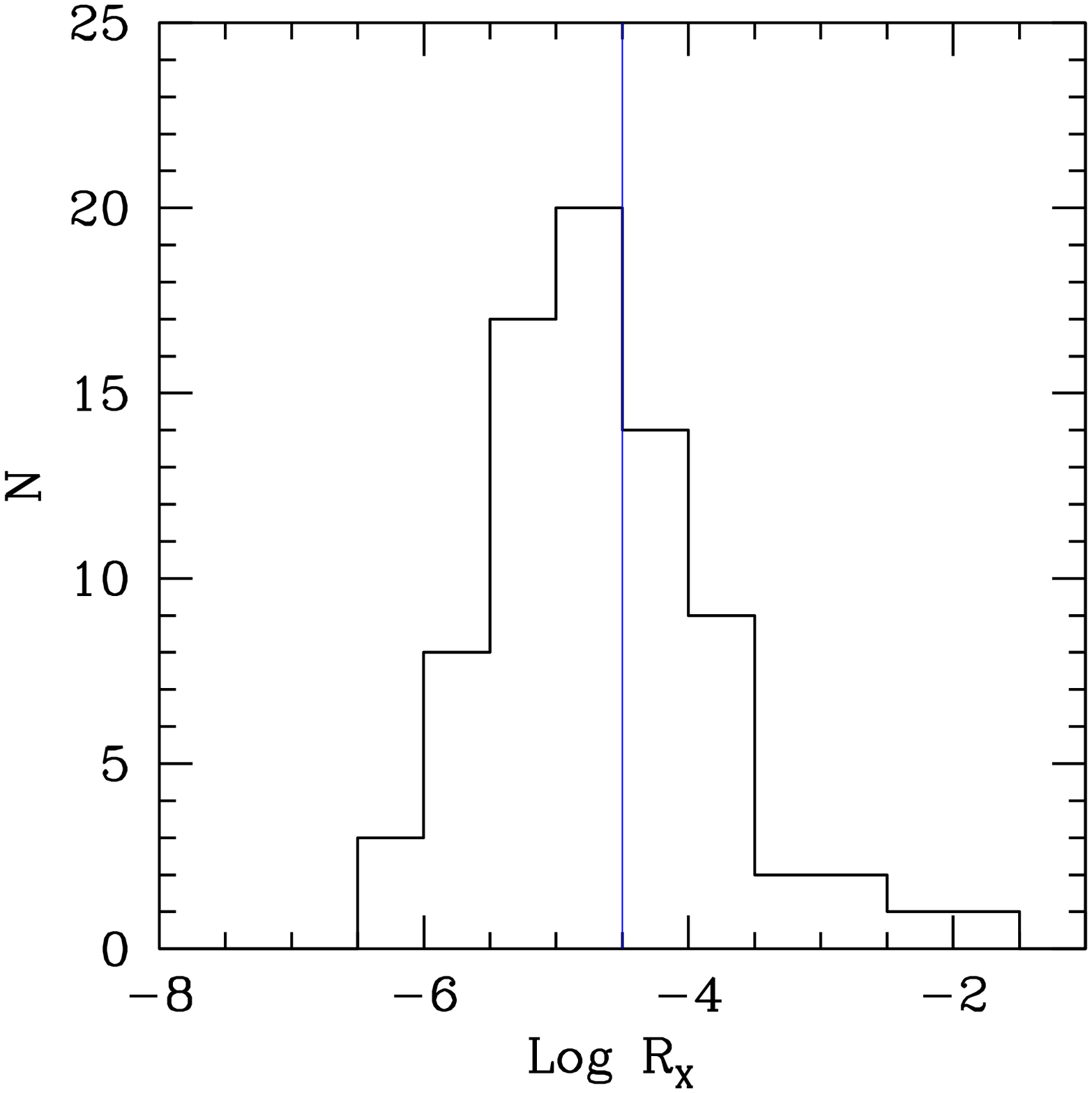}
\includegraphics[width=0.4\textwidth,height=0.3\textheight,angle=0]{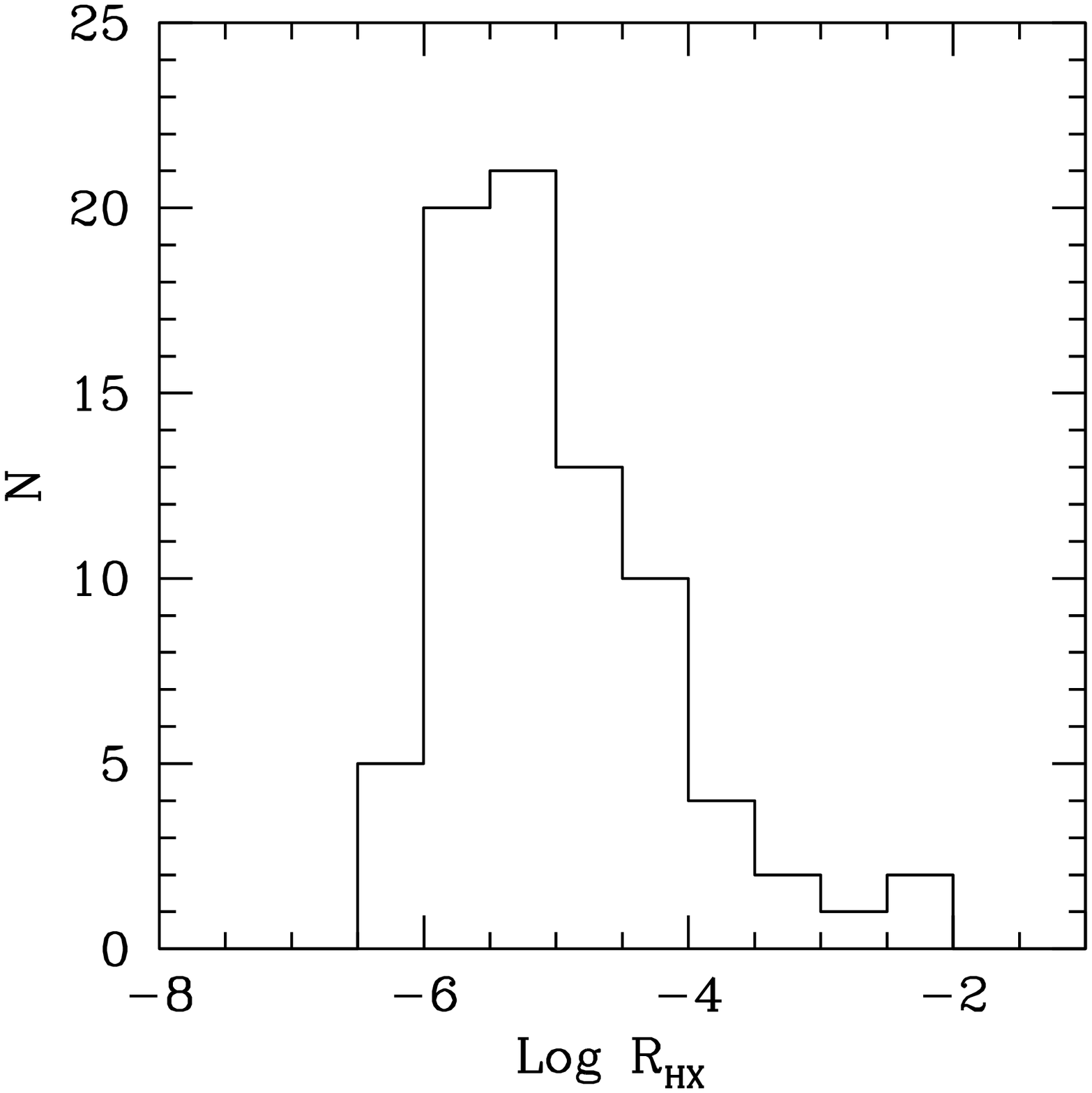}
}
\caption{Left: X-ray Radio loudness given by the ratio of the 1.4 GHz  peak luminosity over the 2-10 keV luminosity. The blue line is the radio-loud versus radio-quiet limit set by Terashima \& Wilson (2003). Right: Hard X-ray Radio loudness given by the ratio of the 1.4 GHz  peak luminosity over the 20-100 keV luminosity.
}
\label{rl}
\end{center}
\end{figure*}

A measure of the radio brightness with respect to other wavelengths is given by the radio loudness, classically
defined by the ratio between the radio and the optical luminosities. Among optically selected AGN, the radio loudness
parameter show a bimodal distribution dividing radio-loud and radio-quiet AGN at the limit of R $\equiv$ L$_{R}$ / L$_{B}$ $\leq$ 10 (Kellermann et al. 1989).
Lately, Terashima \& Wilson (2003) have introduced the X-ray radio-loudness parameter R$_{X}$ $\equiv$ L$_{R}$/L$_{2-10 keV}$. Indeed, the X-ray luminosity is ideal to avoid extinction problems which normally occur in the optical band. 
An even more absorption unbiased measure is given by using the hard X-ray luminosity, therefore
here we define the hard X-ray radio loudness parameter as R$_{HX}$ $\equiv$ L$_{R}$/L$_{20-100 keV}$.
In figure~\ref{rl}, we plot the X-ray and hard X-ray radio-loudness histograms, the blue line shows the radio-loud versus radio-quiet
limit set by Terashima \& Wilson (2003),  i.e. R$_{X}$ $=$ -4.5. The INTEGRAL sample is characterized both by radio-quiet and radio-loud AGN, with a prevalence of radio-quiet AGN,
however the hard X-ray emission is clearly dominant with respect to the radio emission.
From a comparison with the radio-loudness distribution in the BAT sample from Burlon et al. (2013), we find that their sources are more radio-loud (-4 $<$ log R$_{HX}$ $<$ 0).
Again, the high limiting flux density of the AT20G survey allows the detection of only very bright radio sources,
while the use of NVSS/SUMSS data allows the detection of low radio luminosity sources also.
No bimodality is found in the radio-loudness distribution, in agreement with recent findings that
suggest that the bimodality is a consequence of selection biases while well-selected samples of AGN show a continuous level of radio-loudness (e.g., Cirasuolo et al 2003, La Franca et al. 2010).
																						  
In our sample, a wide range of radio morphologies is seen. Thirty-six of the 70 objects (51\%) detected in the NVSS/SUMSS maps show an unresolved (22) or slightly resolved (14) central source coincident, within the uncertainties, with the position of the optical nucleus. Except for the two ``ambiguous'',
the remaining 31 sources (44\% of all detected objects) show evident extended emission, either ``linear'' or ``diffuse'', with
linear sizes ranging from kpc to hundreds of kpc scales. In particular, sources with ``linear'' radio morphology show triple or double emitting structures, thus indicating the presence of large-scale radio jets/lobes, with or without a compact core, respectively, typically 
found in powerful radio galaxies. Indeed, some of the sources in the sample labeled as ``resolved'' (e.g. Cen\,A and Cyg\,A) are among the most famous radio galaxies in the sky.  
Extended-diffuse morphologies might appear unresolved or slightly resolved sources if the host galaxy is distant enough (for the survey angular resolution).			  
However, in our sample of local galaxies, U, S and R sources are equally distributed in redshift; interestingly, one of the most distant sources, IGR~J18249-3243, is one of the  most extended ones (see Landi et al. 2009).

\section{The X-ray versus radio correlation}

\begin{table*}
\begin{center}
\centerline{Table 2: The NVSS data for the INTEGRAL/IBIS complete sample of AGN}
\label{stat}
\begin{tabular}{lcccccccccc}
\hline
\hline
\multicolumn{1}{c}{Variables} &
\multicolumn{1}{c}{N} &
\multicolumn{1}{c}{Spearman} &
\multicolumn{1}{c}{p-value} &
\multicolumn{1}{c}{z-score} &
\multicolumn{1}{c}{p-value} &
\multicolumn{1}{c}{Partial $\tau$} &
\multicolumn{1}{c}{$\sigma$} &
\multicolumn{1}{c}{Prob.} &
\multicolumn{1}{c}{a} &
\multicolumn{1}{c}{b} \\
\multicolumn{1}{c}{(1)} &
\multicolumn{1}{c}{(2)} &
\multicolumn{1}{c}{(3)} &
\multicolumn{1}{c}{(4)} &
\multicolumn{1}{c}{(5)} &
\multicolumn{1}{c}{(6)} &
\multicolumn{1}{c}{(7)}  &
\multicolumn{1}{c}{(8)}  &
\multicolumn{1}{c}{(9)}  &
\multicolumn{1}{c}{(10)}  &
\multicolumn{1}{c}{(11)} \\
\hline
log L$_{1.4 GHz peak}$ vs log L$_{HX}$ & 79 & 0.757 & 0.0        & 7.390 & 0.0	      & 0.33 & 0.05 & 0.0	& 1.21 $\pm$ 0.15  & -14.0   \\
log L$_{1.4 GHz int}$ vs log L$_{HX}$  	 & 70 & 0.673 & 1.1$\times$10$^{-8}$ & 6.136 & 0.0 	      & 0.28 & 0.06 & 0.0	& 1.20 $\pm$ 0.16 & -13.4	  \\
log  L$_{1.4 GHz peak}$ vs log L$_{X}$  	 & 78 & 0.751 & 0.0 & 7.270 & 0.0 	      & 0.32 & 0.06 & 0.0	& 1.17 $\pm$ 0.13 & -12.1    \\
log L$_{1.4 GHz int}$ vs log L$_{X}$ 	 & 70 & 0.658 & 2.3$\times$10$^{-8}$ & 5.918 & 1.7$\times$10$^{-9}$ & 0.28 & 0.06 & 1.4$\times$10$^{-4}$  & 1.08 $\pm$ 0.15 & -7.7      \\
log F$_{1.4 GHz peak}$ vs log F$_{HX}$ 	 & 79 & 0.594 & 7.8$\times$10$^{-8}$ & 5.581 & 1.2$\times$10$^{-8}$ & $-$ & $-$ & $-$	& 1.42 $\pm$ 0.26 & -0.6  \\
log F$_{1.4 GHz peak}$ vs log F$_{X}$	 & 78 & 0.566 & 3.4$\times$10$^{-7}$ & 5.250 & 7.6$\times$10$^{-8}$ & $-$ & $-$ & $-$	& 1.03 $\pm$ 0.20 & -4.3  \\
\hline
\end{tabular}
\end{center}
Notes:  Col. (2-3-4): Number of variables, Spearman's rho correlation coefficient and probability for rejecting the null hypothesis
that there is no correlation; Col. (5-6) Generalized Kendall's $\tau$
z-score and probability for the null hypothesis; Col. (7-8-9):
partial Kendall's $\tau$ correlation coefficient,
the square root of the variance, $\sigma$, and the associated probability P for the null hypothesis; Col. (10-11): Correlation coefficient of the best fit linear 
regression line calculated using Backley-James linear regression method, Y= a $\times$ X + b.
\end{table*}

We have used the {\it ASURV} program (Feigelson \& Nelson 1985, Isobe, Feigelson \& Nelson 1986) for the statistical analysis of our correlations,
which includes several statistical methods that deal with data sets containing both detections and non-detections (upper limits in our case).
Within the program, we have used the generalized version of Spearman's rank order correlation coefficient and the generalized Kendall's $\tau$ test
to compute the correlation significance. The Buckley-James regression method has been used to calculate the linear regression coefficients.
In table 2 we report the results of the statistical analysis.

\begin{figure*}
\begin{center}
\parbox{16cm}{
\includegraphics[width=0.4\textwidth,height=0.3\textheight,angle=0]{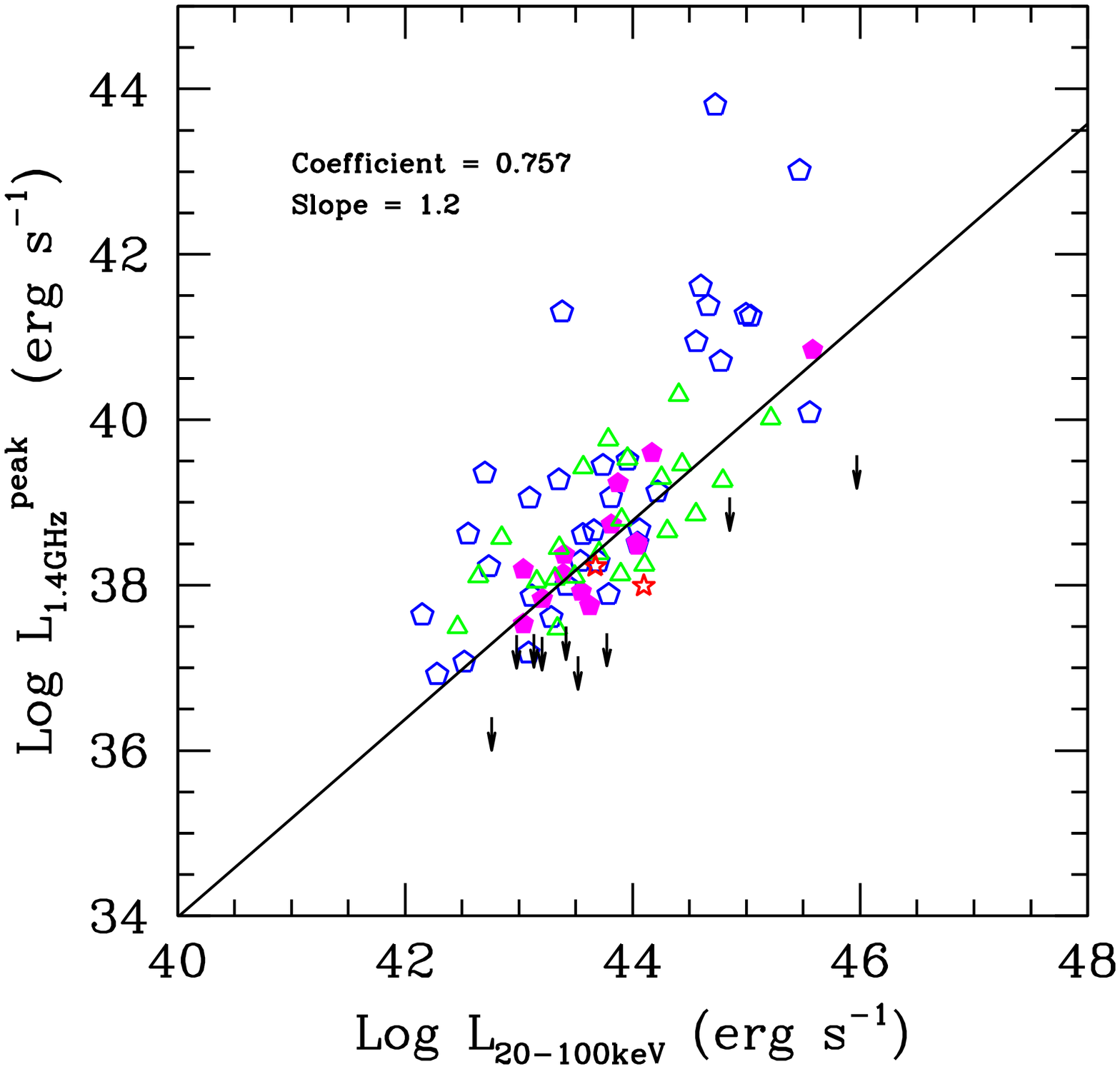}
\includegraphics[width=0.4\textwidth,height=0.3\textheight,angle=0]{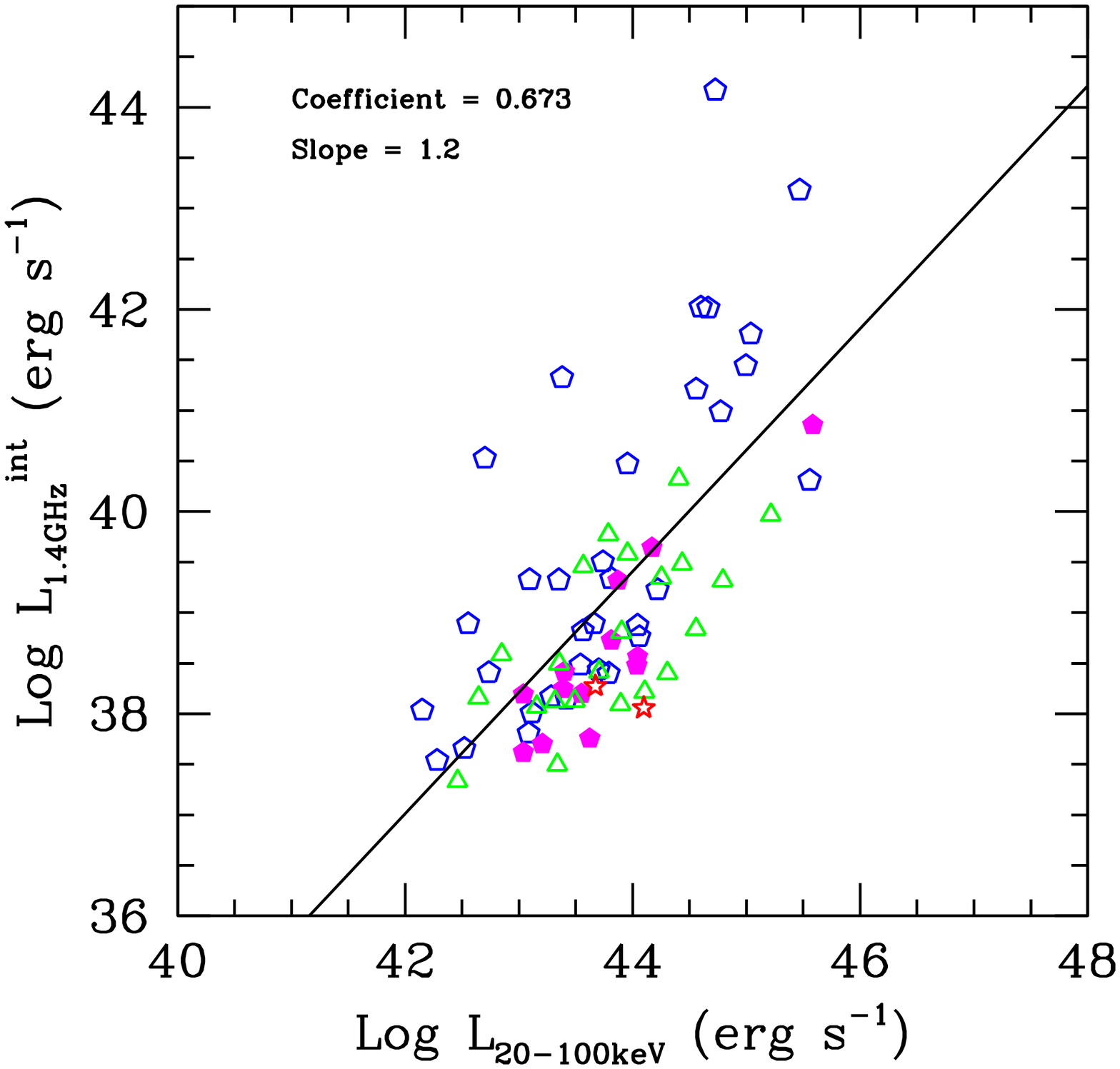}}
\parbox{16cm}{
\includegraphics[width=0.4\textwidth,height=0.3\textheight,angle=0]{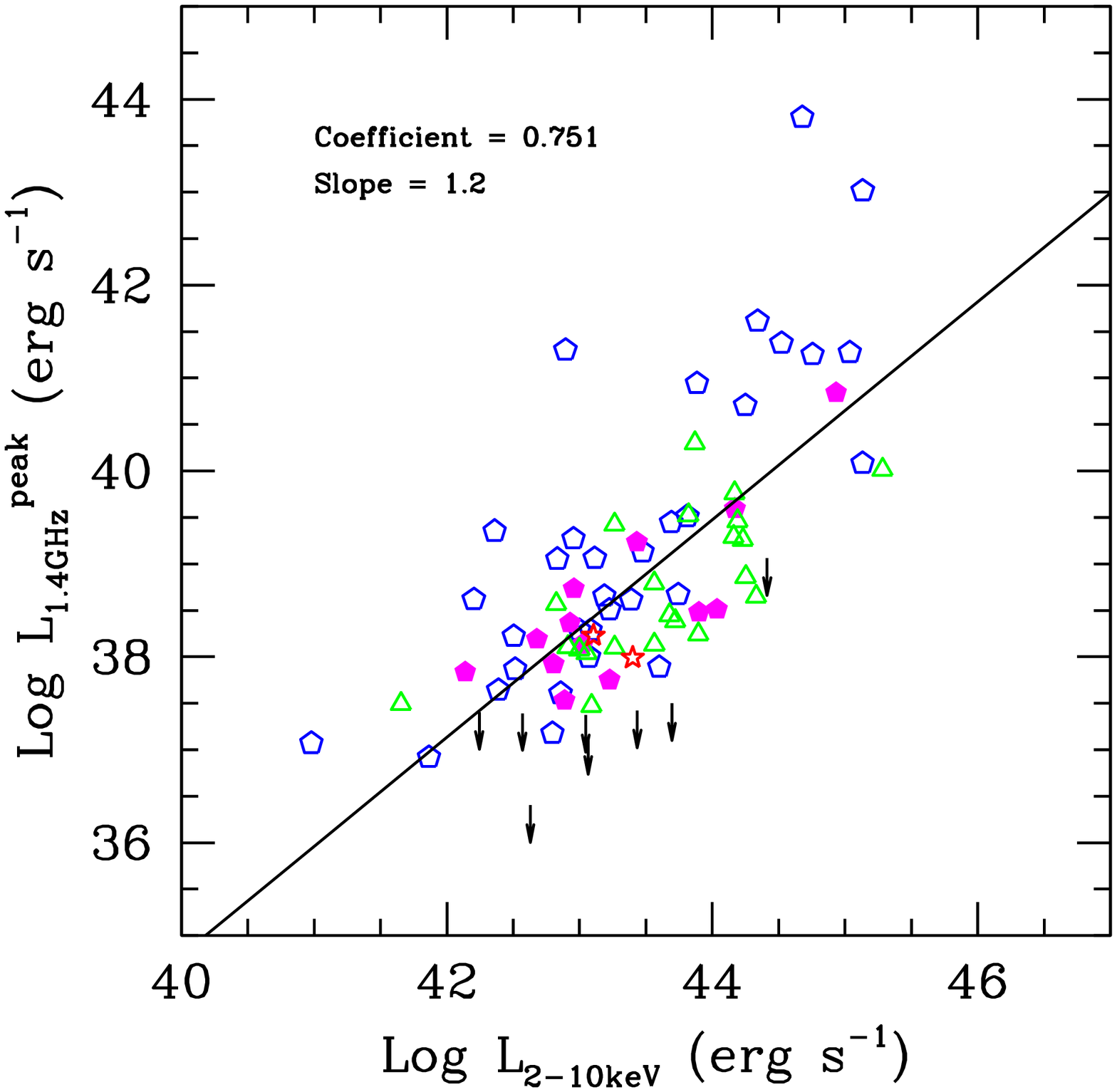}
\includegraphics[width=0.4\textwidth,height=0.3\textheight,angle=0]{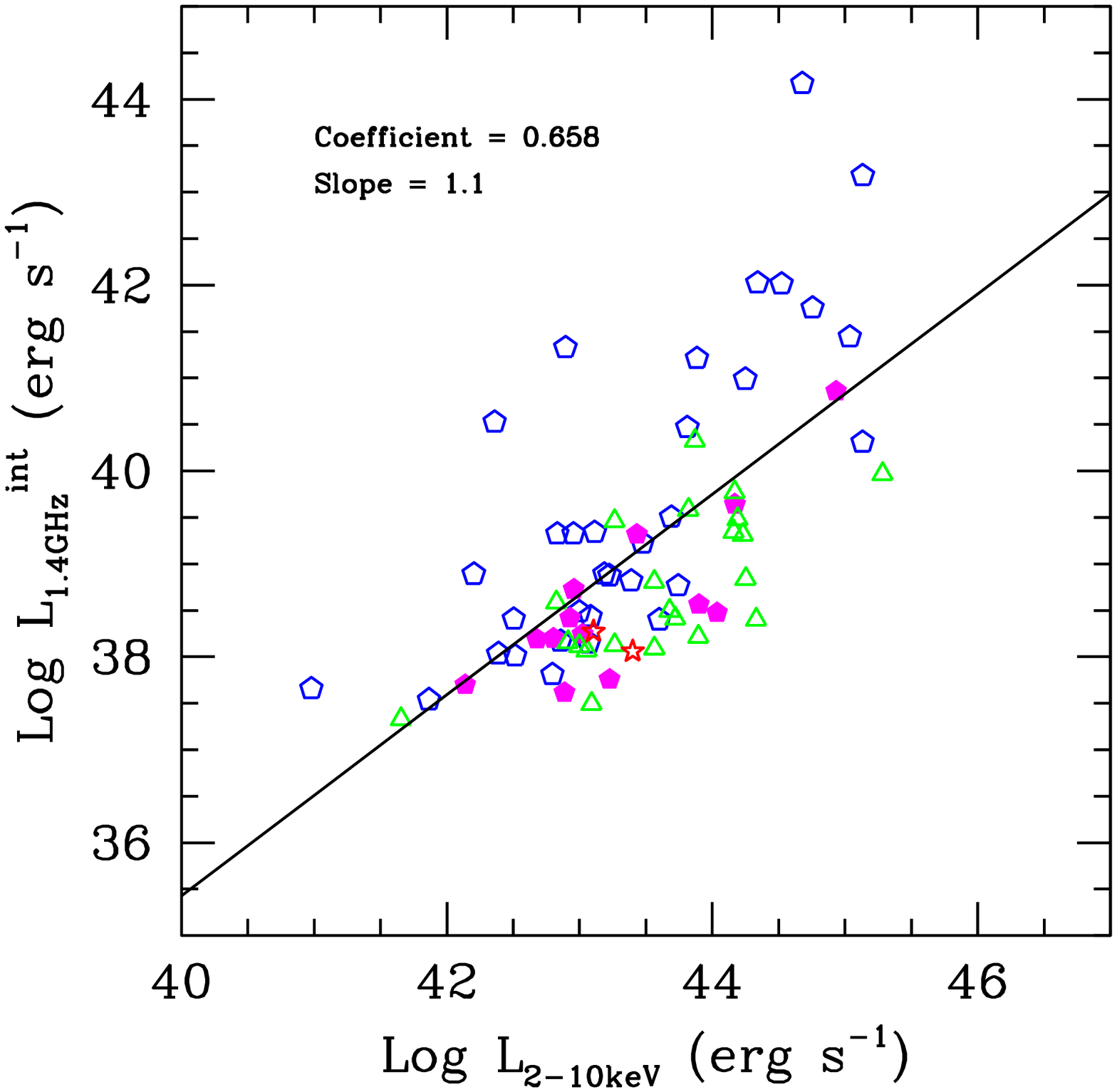}}
\parbox{16cm}{
\includegraphics[width=0.4\textwidth,height=0.3\textheight,angle=0]{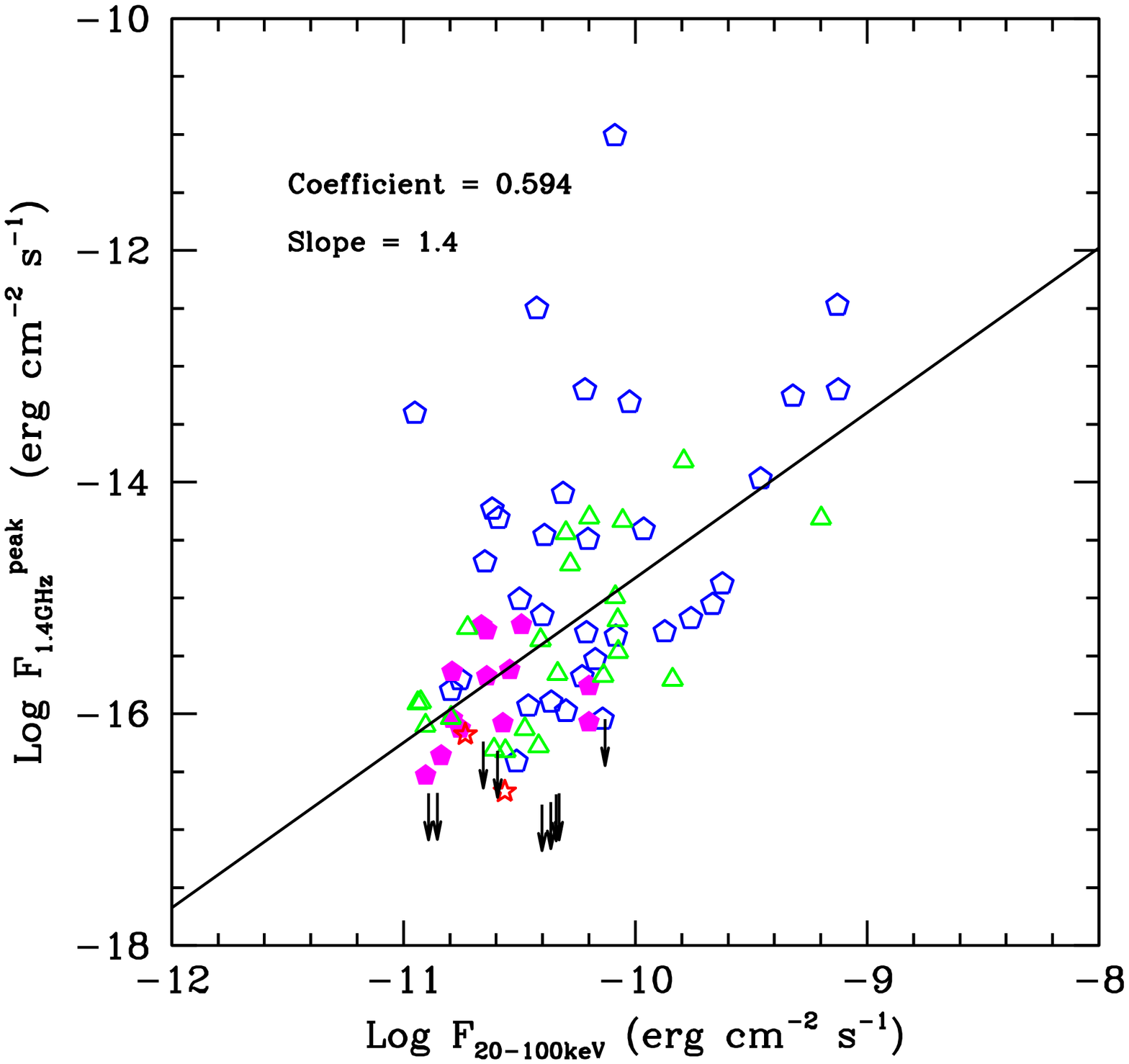}
\includegraphics[width=0.4\textwidth,height=0.3\textheight,angle=0]{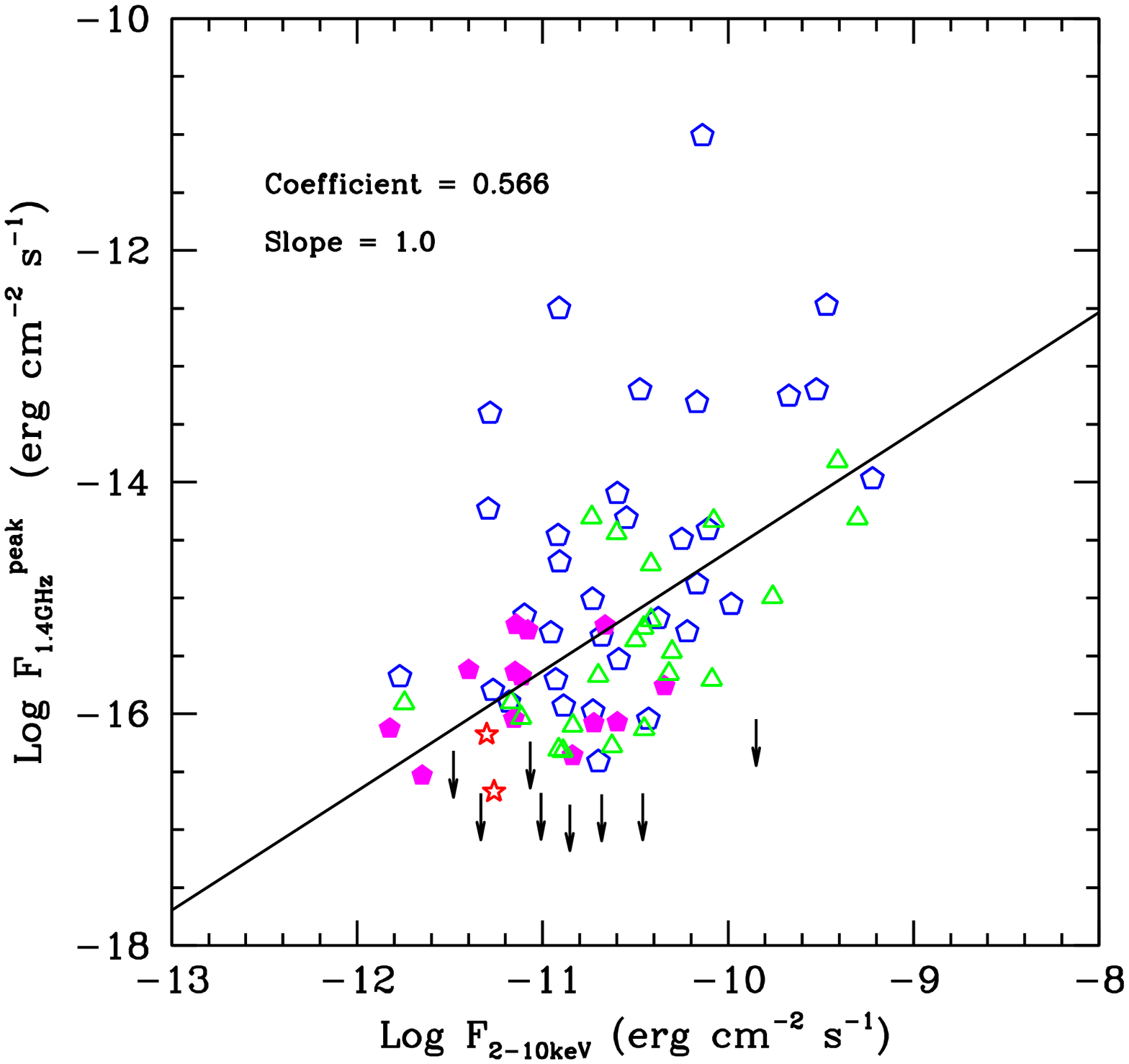}}
\caption{Upper panels: 1.4 GHz peak (left) and integrated (right) versus 20-100 keV luminosity. Middle panels: 
1.4 GHz peak (left) and integrated (right) versus 2-10 keV luminosity. Bottom panels: 1.4 GHz  peak versus 20-100 keV (left) and 2-10 (right) fluxes. 
The relative best fit regression line is shown for each correlation. Different morphological classes are marked as follows: resolved sources (R)
are blue empty pentagons, slightly resolved (S) sources are magenta solid pentagons, unresolved (U) sources are green empty triangles, ambiguous (A) sources are red stars. Upper limits are marked as black arrows.}
\label{fig:corr}
\end{center}
\end{figure*}

In figure~\ref{fig:corr} we show the 1.4 GHz peak (left) and integrated (right) luminosity versus the 20-100 keV luminosity with its best fit regression lines.
The two correlations are both highly significant (see table 2) 
and display a very similar slope (1.21 $\pm$ 0.15 and 1.20 $\pm$ 0.16, respectively).
A similar result is obtained if the radio peak and integrated luminosities are plotted against the 2-10 keV luminosity (figures~\ref{fig:corr}),
i.e., high significance and correlation slopes of 1.12 $\pm$ 0.13 and 1.08 $\pm$ 0.15, consistent within errors.
We note that hard X-ray fluxes show typically larger values than the 2-10 keV ones, this naturally yields to a steeper correlation slope.

Luminosity-luminosity correlations may be distance driven. To check for a possible dependence on the distance we have 
performed a partial Kendall $\tau$ correlation test, which computes partial correlation 
coefficient and significance for censored data using three variables, where we took the distance as the third variable (Akritas 
\& Siebert 1996). The partial KendallÕs $\tau$ test results (see table 2) suggest that the correlations are not driven 
by distance effects. As a further check we compute the correlation significance for the flux-flux plot of 1.4~GHz peak flux versus 20-100 keV and 2-10 keV fluxes (figure~\ref{fig:corr}) and, despite the large scatter, the correlation remains statistically signiÞcant.

We compare our results with those from Burlon et al. (2013), in which the 15-55 keV BAT luminosities are
correlated with the 20 GHz ones. As pointed out by the authors, at 20 GHz most of the extended structures are resolved, 
allowing a more direct comparison between the core radio emission and the hard X-rays. 
The correlation slope for their Seyfert sample is log L$_{15-55 keV}$ $\propto$ 0.7 log L$_{20 GHz}$, slightly steeper but in agreement within errors with the values found in our sample;
however, the weak correlation found is clearly driven by distance effects, in fact their sample is characterized by several sources at high z.

\begin{figure*}
\begin{center}
\includegraphics[width=0.4\textwidth,height=0.3\textheight,angle=0]{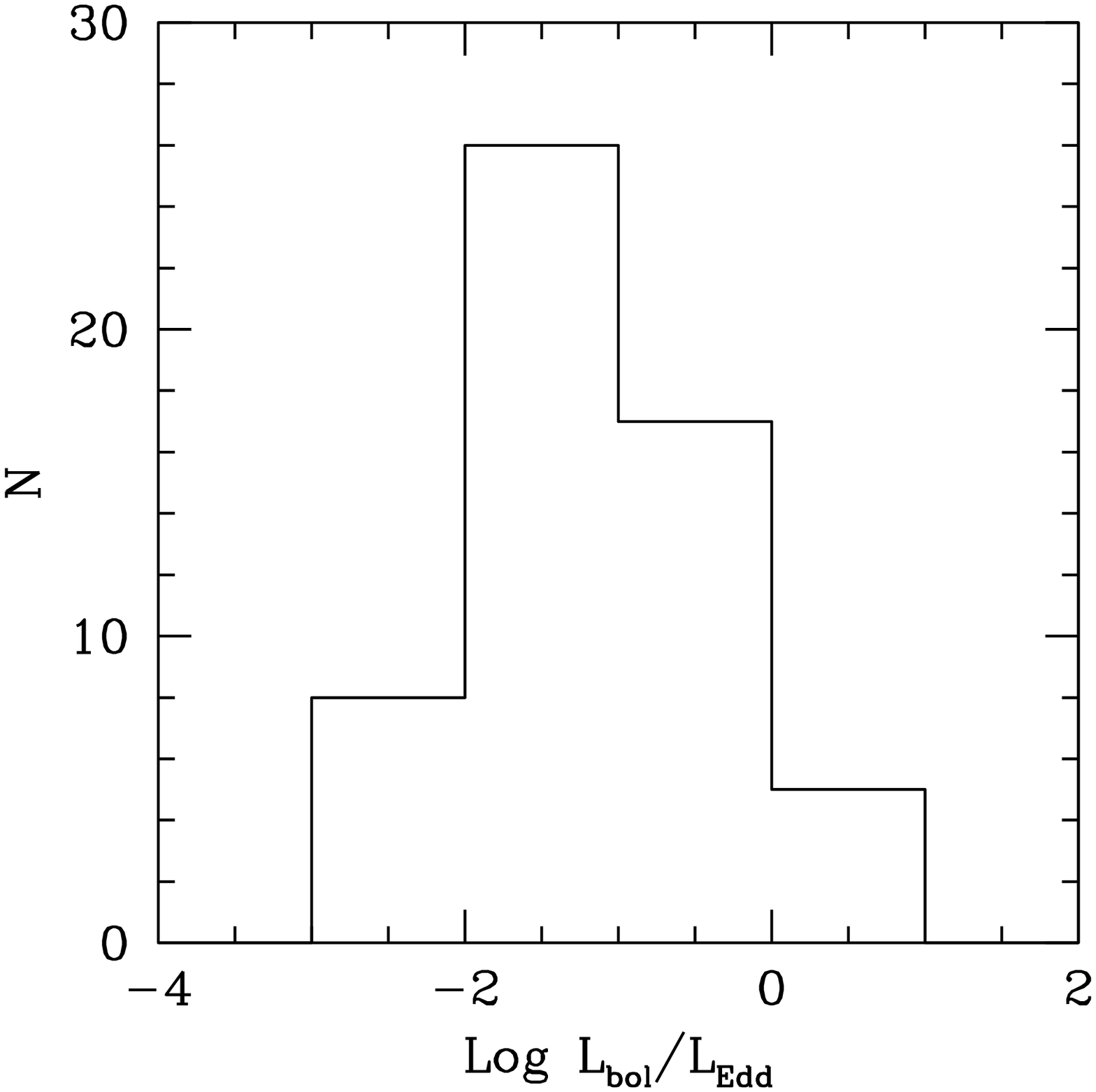}
\caption{Distribution of the Log Eddington ratio defined as the logarithm of the ratio between the bolometric luminosity and the Eddington luminosity,  L$_{bol}$/L$_{Edd}$.}
\label{bol}
\end{center}
\end{figure*}

\section{Discussion}

In black hole systems, the X-ray emission typically originates in the accretion flow surrounded by a hot corona or alternatively at the base of a jet (Markoff et al. 2005). The radio emission is associated with the presence of a jet of relativistic particles emitting synchrotron radiation at radio frequencies. AGN with powerful relativistic jets only makes up 10-20\% of the entire AGN population (Kellermann et al. 1989), the majority of which is instead quiet in the radio, although not silent (Ho \& Ulvestad 2001). Radio cores are almost ubiquitously observed depending on the survey flux limit and frequency. The 89\% detection rate of radio sources found within the INTEGRAL AGN survey is in agreement with these previous statements.

The existence of a relation between the X-ray and the radio emission in both stellar and super-massive black holes has opened new scenarios on the comprehension of the accretion-ejection mechanisms (Corbel et al. 2003, Gallo et al. 2003, Merloni et al. 2003, Falcke et al. 2004).  Stellar-mass black holes follow a standard relation L$_{R}$ $\propto$ L$_{X}^{0.5-0.7}$ when they are found in their soft state (Corbel et al. 2003, Gallo et al. 2003). Interestingly, the fundamental plane, which unifies XRBs and AGN by scaling the radio and X-ray luminosity with the black hole mass, displays the same slope (Merloni et al. 2003). For these classes of sources it has been proposed that an inefficient accretion flow model coupled with a (scale invariant) jet can reproduce the observed slope. Indeed, when the disk accretes at radiatively inefficient regime, the advected flow plus jet can explain both the X-ray and radio emission (e.g., Narayan \& Yi 1994). 

Since the discovery of the original correlation, more and more outliers to the standard correlation are found, i.e., for a given X-ray luminosity the radio emission is fainter resulting in steeper slopes (e.g., Soleri et al. 2010). Coriat et al. (2011) propose that, assuming a standard emission model of compact jets, the steeper relation ($\sim 1.4$) implies that the accretion flow is radiatively efficient, where the limit for the transition between radiatively inefficient to radiatively efficient is set to an X-ray luminosity higher than 5 $\times$ 10$^{-3}$ L$_{Edd}$.
Dong et al. (2014) have confirmed this hypothesis for a sample of high Eddington ratio AGN,  by assuming  L$_{R}$ $\propto$ Q$_{jet}^{1.4}$ $\propto$ $\dot{M}^{1.4}$.
The significant correlation found for the INTEGRAL AGN sample confirms these previous works.
The 1.4 GHz versus 20-100 keV emission correlates with a slope of 1.2$\pm$0.15, steeper than the classical 0.5-0.7, and
consistent within errors with the expected 1.4. Indeed, the Eddington ratios (L$_{bol}$/L$_{Edd}$ $\geq$ 10$^{-3}$, where the bolometric luminosity is derived by applying a correction factor of 20 to the 2-10 keV luminosity, see Vasudevan \& Fabian (2007)) of the INTEGRAL sample (see figure~\ref{bol}), strongly suggest that these sources are efficient accreting systems. 
as expected from their hard X-ray selection and behave like the outliers XRB systems discussed by Coriat et al.  (2011).
All the above theoretical considerations assume a standard behavior for the jet and ascribe the observed correlation slope 
to a particular accretion regime, however the jet emission depends on several parameters, a change in one of those (e.g., the magnetic field)
could modify the radio output with strong influence in the correlation slope (see discussion in Coriat et al. 2011).

We note that the radio data here presented comprise the emission over kpc scales extension.
At intermediate radio resolutions (such as those of the VLA-A observation, i.e. arcsec), the emission comes from pc scales and 
the X-ray emission still correlates with the radio emission at 1.4, 5 and 15 GHz (Panessa et al. 2007), with a similar level of significance. 
However, at VLBI scales (milli arcsec), no correlation is found  for a sample of Seyfert galaxies, suggesting that the very inner radio core emission at sub-pc scales is
not related to the X-rays (Panessa \& Giroletti 2013). All these evidences support the idea that the disk-corona emission correlates with the total radio emission, rather than
with the inner jet emission.

However, all the above mentioned works cannot rely on the simultaneity of the radio versus X-ray data. When simultaneous data are taken, only very weak correlations are found
(King et al. 2011, 2013, Jones et al. 2011). For both nearby AGN NGC~4051 and NGC~4395, the intrinsic X-ray variability is observed to be stronger than the radio variability. Assuming that
the corona is at the base of a jet, the X-ray variations may be filtered by the corona and only partially transferred to the jet. However in these works, radio data are at arcsec scales,
implying that the radio emission is averaged on a few hundreds pc, the different X-ray versus radio spatial scales sampled could explain the different variability.
Interestingly the simultaneous L$_{R}$ $\propto$ L$_{X}$ slopes are steeper (around unity) than the standard 0.6, suggesting that these sources are X-ray bright and that their emission is accretion-dominated.

In interpreting the X-rays and hard X-rays versus 1.4 GHz correlations with our data sets we should consider two more important aspects.
One is that the lower the radio frequency is, the more the emission is contaminated by diffuse components, such as lobes on a few kpc to hundreds of kpc spatial scales.
This implies that the time scales sampled by the NVSS and SUMSS surveys varies over thousands of years, i.e. the time needed by electrons to propagate into the medium.
On the other hand, the X-ray flux is averaged over tens of ksec observations (few days), while the INTEGRAL hard X-ray emission is averaged over months to several years of sky surveying.
Both the X-ray and hard X-ray emission are produced within the inner region of the accretion flow (inner disk - corona), therefore the X-ray spatial scale sampled is of a few gravitational radii.
Summarizing, the radio emission is averaged on much larger spatial and time scales with respect to the X-ray and hard X-ray emission. As a consequence, the present correlations suggest that the 
accretion related X-ray emission is related to the diffuse/total radio emission emitted over larger scales.
Within our group we are developing a model which correlates the hard X-ray emission with the total radio emission assuming a more or less spherical bubble on scales of several arc minutes, 
fed by jets issuing from the core (see Wagner et al. 2012, Wagner \& Bicknell 2011). Details on this theoretical treatment and interpretation of the X-ray/radio correlations examined here and its implications will be presented and discussed in a future work (Bicknell et al. in preparation).

\section{Conclusion}

We have selected a complete sample of hard X-ray AGN from the INTEGRAL survey in order to study the relation between
the X-ray and hard X-ray emission with the radio emission in a sample of bright sources.
First we have analyzed the radio images from the NVSS and SUMSS surveys detecting a large fraction of sources (89\%),
with different morphologies, from unresolved/slightly resolved to extended sources over hundreds of kpc scales.
We find the existence of a significant correlation between the X-ray and hard X-ray luminosities and the peak and integrated 1.4 GHz radio luminosities. Correlation slopes are around 1-1.2, 
clearly steeper than the classical 0.6 value found in the fundamental plane for BH activity and 
consistent with the 1.4 expected for sources belonging to the efficient accretion branch (Coriat et al. 2011, Dong et al. 2014).
This suggests that the INTEGRAL AGN are X-ray dominated (accretion dominated) sources accreting at high Eddington ratios
and the high energy emission from the central engine is related to the radio emission averaged over kpc scales (i.e.,  thousands of years).\\

{\bf Acknowledgements}\\
We acknowledge the anonymous referee for his/her suggestions. 
FP thanks Davide Burlon and Piergiorgio Casella for the fruitful scientific discussions. AT and PC would like to thank Matteo Murgia for useful suggestions on the analysis of radio data.
EM acknowledges Paola Parma for the technical support during part of the radio data analysis.
F.P. acknowledges support by INTEGRAL ASI/INAF n. 2013-025.R.O.

\end{document}